\documentclass[preprint,authoryear,12pt]{elsarticle}

\makeatletter
\def\ps@pprintTitle{%
 \let\@oddhead\@empty
 \let\@evenhead\@empty
 \def\@oddfoot{\centerline{\thepage}}%
 \let\@evenfoot\@oddfoot}
\makeatother

\usepackage{graphicx}
\usepackage{epstopdf}
\usepackage{booktabs}
\usepackage{natbib} 
\usepackage{multirow}
\usepackage{amssymb}
 \usepackage{amsthm}
 \usepackage{amsmath}
 \usepackage{lscape}
 \usepackage{stmaryrd}

\journal{Springer Proceedings in Complexity}

\begin{document}

\begin{frontmatter}

\title{Power-law cross-correlations: Issues, solutions and future challenges}

\author[utia]{Ladislav Kristoufek} \ead{kristouf@utia.cas.cz}

\address[utia]{Institute of Information Theory and Automation, Czech Academy of Sciences, Pod Vodarenskou vezi 4, Prague, Czech Republic, EU}

\begin{abstract}
Analysis of long-range dependence in financial time series was one of the initial steps of econophysics into the domain of mainstream finance and financial economics in the 1990s. Since then, many different financial series have been analyzed using the methods standardly used outside of finance to deliver some important stylized facts of the financial markets. In the late 2000s, these methods have started being generalized to bivariate settings so that the relationship between two series could be examined in more detail. It was then only a single step from bivariate long-range dependence towards scale-specific correlations and regressions as well as power-law coherency as a unique relationship between power-law correlated series. Such rapid development in the field has brought some issues and challenges that need further discussion and attention. We shortly review the development and historical steps from long-range dependence to bivariate generalizations and connected methods, focus on its technical aspects and discuss problematic parts and challenges for future directions in this specific subfield of econophysics.
\end{abstract}


\end{frontmatter}

\newpage

\section{Introduction}

Analysis of long-range dependence properties of financial time series was at the very beginning of the econophysics field in the early 1990s \citep{Beran1994,Mantegna2000,Samorodnitsky2006} following the early works of the Mandelbrot research group \citep{Mandelbrot1967,Mandelbrot1968,Mandelbrot1968a}. At the time, most of the financial works were based on the assumption that, in addition to other simplifying restrictions, the auto-correlation function of the series vanishes exponentially, i.e. very quickly. Lagged observations of the series thus played only a marginal role and only after few time steps, the effect was assumed to be gone completely. Such assumption has some convenient mathematical properties in a parallel logic to assuming the Gaussian distribution. However, the noted works, among others, have argued that some financial time series show that observations at even very high lags can have an effect on current movements of the financial series. This gave rise to the so-called Hurst effect with respect to \cite{Hurst1951} -- and his work in hydrology -- which has since been referred to by various names, mostly persistence, long-range dependence, and long-term correlations (and sometimes long-term memory).

Long-range dependence of time series is characteristic by a slowly decaying auto-correlation function, contrary to the quickly vanishing exponentially decreasing auto-correlation function standardly seen in autoregressive (integrated) moving-average processes (ARMA/ARIMA) \citep{Box1994} and (generalized) autoregressive conditional heteroskedasticity models (ARCH/GARCH) \citep{Engle1982,Bollerslev1986} that are standard in financial economics and financial econometrics. In the econophysics literature, the slowly decaying auto-correlation function is usually represented by a hyperbolical decay. Such specification has some intriguing properties \citep{Beran1994,Samorodnitsky2006} which allowed for introduction of many estimators of long-range dependence parameters, but most importantly to the detrended fluctuation analysis (DFA) \citep{Peng1993,Peng1994}, which quickly became the most popular method of studying long-range dependence properties in the time domain. Its simplicity and intuitive appeal made it an ideal candidate for various specifications, adjustments and generalizations -- most notably the multifractal detrended fluctiation analysis (MF-DFA) \citep{Kantelhardt2002} and detrended cross-correlation analysis (DCCA) \citep{Podobnik2008}. The former method generalizes the original one by studying multifractal properties rather than (mono/uni)fractal ones and the latter studies the long-range dependence properties between two series, i.e. cross-correlations rather than serial (auto-)correlations.

In this work, we study and review the methodological steps that needed to be taken when coming from long-range correlations towards long-range cross-correlations. Importantly, we focus on problematic parts of the latter and cover two approaches how to treat them. Specifically, we argue (and review the relevant literature that shows so) that long-range (power-law) cross-correlations are only an in-between step and by themselves, they tell very little. The two approaches, which utilize the power-law cross-correlations as the mentioned in-between step, are the scale-specific correlations and regressions, and the power-law coherency. Eventually, we show that these two are inherently related. In our discussion, we outline possible future challenges in this branch of interdisciplinary research.

\section{From long-range dependence to power-law cross-correlations}

Persistent series can be characterized through its dynamic properties in both time and frequency domains. In the former, the auto-correlation function is standardly represented by an asymptotic hyperbolic decay, specifically $\rho(k) \propto k^{2H-2}$ for $k \rightarrow +\infty$ where $\rho(k)$ is the auto-correlation function at time lag $k$ and $H$ is the Hurst exponent \citep{Samorodnitsky2006}. In the latter, the persistence translates into a power-law divergence of the spectrum at origin, specifically $f(\omega)\propto \omega^{1-2H}$ for $\omega \rightarrow 0+$ where $f(\omega)$ is the spectrum and $\omega$ is the frequency \citep{Beran1994}. The critical parameter here is the Hurst exponent $H$. For the stationary series, the exponent ranges between 0 and 1 and is well separated by 0.5 which marks a process with no persistence. Processes with $H>0.5$ are the persistent ones that have strong auto-correlation structure and remind of locally trending processes that still remain stationary. Anti-persistent processes with $H<0.5$ are characteristic by frequent switching of signs of their changes but are usually of a marginal interest compared to the persistent processes that can be exploited in finance for profitable trading strategies \citep{Mandelbrot1968a}. Even though the frequency domain approach has gained more traction in the financial econometrics field, it has been the time domain estimators that became more popular in the interdisciplinary research. We follow this logic and focus on the the time domain implications of long-range dependence (even though most of it can be quite easily translated into the frequency domain language).

The hyperbolic decay of the auto-correlation function has some interesting implications which are covered in various textbooks (we refer here to the ``classics'' of \cite{Beran1994} and \cite{Samorodnitsky2006}) but specifically its connection to the scaling of partial sums has crucial application. We define a partial sum of process $\{x_t\}_{t=1}^T$, where $T$ is the time series length, as $X_t=\sum_{i=1}^t{x_i}$. If $\{x_t\}_{t=1}^T$ is long-range correlated, then the variance of partial sums scales as $\text{Var}(X_t) \propto t^{2H}$ for $t \rightarrow +\infty$. It turns out that variance of an integrated process (the partial sum) is much less noisy than auto-correlation function of the original process at high lags, which in turn makes the approach based on the partial sums and variance more appropriate for estimation of the Hurst exponent $H$.

The partial sums divergence is utilized in various estimators of the Hurst exponent, most notably by the detrended fluctuation analysis (DFA)\citep{Peng1993,Peng1994,Kantelhardt2002}. DFA is based on several steps mainly focused on further reducing the noise in the estimation procedure as well as filtering out possible time trends. Specifically, one starts with a profile of the series (a cumulative sum of the de-meaned original series), which represents the cumulative sum in the previous paragraph. Such profile is split into intervals of length $s$ representing a scale. In each interval of the given length, a time trend is estimated (usually a linear trend but the procedure can utilize many different filtering procedures) and a mean squared error around the trend is found. This squared error is then averaged over all intervals of the given length to give a fluctuation function $F^2(s)$. The procedure is repeated for a range of scales and the Hurst exponent is estimated on the scaling rule $F^2(s) \propto s^{2H}$. DFA has become and remained the most popular of the time domain Hurst exponent estimators even over its weaknesses as reported in various studies \citep{Taqqu1995,Taqqu1996,Teverovsky1999,Grech2005,Barunik2010,Kristoufek2010} mainly due its straightforward nature and implementation. Although, it needs to be stressed that DFA is also the most tested and numerically examined of the methods.

It took more than a decade to come from DFA to a parallel examination of dependence between two series. And again, it was DFA in the center. \cite{Podobnik2008} introduced the detrended cross-correlation analysis (DCCA/DXA) that is built on a parallel idea -- scaling of covariances between partial sums. Even though the step from DFA to DCCA is intuitively clear and frankly trivial -- instead of finding a mean squared error from the trend in each window of size $s$, we find a product of errors from the trend for two series -- it took five more years to prove that in fact the covariance of partial sums of two series, out of which at least one is long-range correlated, scales as $\text{Cov}(X_t,Y_t) \propto t^{2H_{xy}}$ where $H_{xy}$ is the bivariate Hurst exponent \citep{Kristoufek2013}. And even though DCCA has quickly become popular and widely used in the empirical literature across disciplines, the theoretical paper of \cite{Kristoufek2013} was one of the first to show that going from univariate to bivariate perspective has some serious methodological caveats and making intuitive translations from the former to the latter without proper theoretical treatment can lead to crucial errors.

\section{The issues with power-law cross-correlations}

Most of the literature building on the DCCA procedure has been empirical and it has become quickly clear that the relationship between the bivariate Hurst exponent $H_{xy}$ and the Hurst exponents of the separate processes $H_x$ and $H_y$ might play a crucial role. From one side, most empirical studies reported that either $H_{xy}=\frac{1}{2}(H_x+H_y)$ or $H_{xy}>\frac{1}{2}(H_x+H_y)$ \citep{He2011,Wang2013,Oswiecimka2014}. From the other, numerical and theoretical studies suggested that either $H_{xy}=\frac{1}{2}(H_x+H_y)$ or $H_{xy}<\frac{1}{2}(H_x+H_y)$ \citep{Sela2012}. The clash was apparent and a more detailed theoretical treatment was clearly needed.

The primary issue of the literature (both theoretical and empirical) on power-law cross-correlations was non-existence of a process that would generate power-law cross-correlated series and allow to control the $H_{xy}$ parameter. Even though \cite{Podobnik2008a} proposed the two-component ARFIMA process as a mixture of two power-law auto-correlated processes, it has not been numerically shown how to control the $H_{xy}$ parameter as a function of $H_x$, $H_y$ and the proposed weight $W$. The proposed process was verified by the DCCA estimation, which, unfortunately, is not a proper way of proving validity as DCCA itself has not been shown to have clear statistical properties. This circular proof is thus not valid.

\cite{Kristoufek2013b} introduced the mixed-correlated ARFIMA process (MC-ARFIMA), which allowed for controlling the $H_{xy}$ parameter. MC-ARFIMA processes are defined as

\begin{gather}
x_t=\alpha\sum_{n=0}^{+\infty}{a_n(d_1)\varepsilon_{1,t-n}}+\beta\sum_{n=0}^{+\infty}{a_n(d_2)\varepsilon_{2,t-n}} \nonumber \\
y_t=\gamma\sum_{n=0}^{+\infty}{a_n(d_3)\varepsilon_{3,t-n}}+\delta\sum_{n=0}^{+\infty}{a_n(d_4)\varepsilon_{4,t-n}} \nonumber
\label{eq:ARFIMA_LC}
\end{gather}
where 
\begin{equation}
a_n(d)=\frac{\Gamma(n+d)}{\Gamma(n+1)\Gamma(d)} \nonumber
\end{equation}
and error terms are characterized by
\begin{gather}
\langle \varepsilon_{i,t} \rangle = 0\text{ for }i=1,2,3,4 \nonumber\\
\langle \varepsilon_{i,t}^2 \rangle = \sigma_{\varepsilon_i}^2\text{ for }i=1,2,3,4 \nonumber\\
 \langle \varepsilon_{i,t}\varepsilon_{j,t-n} \rangle = 0\text{ for }n \ne 0\text{ and }i,j=1,2,3,4 \nonumber\\
\langle \varepsilon_{i,t}\varepsilon_{j,t} \rangle = \sigma_{ij}\text{ for }i,j=1,2,3,4\text{ and }i\ne j. \nonumber
\end{gather}

To put it in words, the two processes are each a linear combination of two power-law auto-correlated processes with possibly correlated error-terms. The separate long-term memory parameters $d_1,d_2,d_3,d_4$ are unrestricted. The $d$-notation is kept here mainly due to the use of the $\Gamma(\bullet)$ function and it standardly holds that $H=d+\frac{1}{2}$. Even though the paper discusses more possibilities, there are two important specifications. First, if we do not restrict the correlation between error-terms in any way, the bivariate Hurst exponent will be an average of the separate Hurst exponents. And second, if the two processes with lower separate Hurst exponents in each have correlated error-terms and the other error-terms are uncorrelated, the bivariate Hurst exponent will be lower than the average of the two separate ones. There is no combination of parameters that would allow the bivariate Hurst exponent to be higher than the average of the separate ones. This can be quite easily seen from the asymptotic behavior of the cross-correlation function between two MC-ARFIMA processes (and more details are given in the reference):
\begin{multline}
\rho_{xy}(n)=\ldots \approx \\
\frac{\alpha\gamma\sigma_{13}}{\sigma_x\sigma_y}\underbrace{\sum_{k=0}^{+\infty}{a_k(d_1)a_{n+k}(d_3)}}_{\approx \int_{0}^{+\infty}{k^{d_1-1}(n+k)^{d_3-1}dk}\propto n^{d_1+d_3-1}}+\frac{\alpha\delta\sigma_{14}}{\sigma_x\sigma_y}\underbrace{\sum_{k=0}^{+\infty}{a_k(d_1)a_{n+k}(d_4)}}_{\approx \int_{0}^{+\infty}{k^{d_1-1}(n+k)^{d_4-1}dk}\propto n^{d_1+d_4-1}}+\\
\frac{\beta\gamma\sigma_{23}}{\sigma_x\sigma_y}\underbrace{\sum_{k=0}^{+\infty}{a_k(d_2)a_{n+k}(d_3)}}_{\approx \int_{0}^{+\infty}{k^{d_2-1}(n+k)^{d_3-1}dk}\propto n^{d_2+d_3-1}}+\frac{\beta\delta\sigma_{24}}{\sigma_x\sigma_y}\underbrace{\sum_{k=0}^{+\infty}{a_k(d_2)a_{n+k}(d_4)}}_{\approx \int_{0}^{+\infty}{k^{d_2-1}(n+k)^{d_4-1}dk}\propto n^{d_2+d_4-1}}. \nonumber
\label{eq:rhon_MC-ARFIMA}
\end{multline}

The MC-ARFIMA introduction has had two main results. First, there was finally a data generator that could be used for simulation studies that also has well-defined statistical properties \citep{Kristoufek2015a,Kristoufek2016}. And second, the possibility of having $H_{xy}>\frac{1}{2}(H_x+H_y)$ seemed to have vanished as the MC-ARFIMA processes are very generally defined and allow for very flexible manipulation. In other words, if it was not possible to find a specification that would lead to $H_{xy}>\frac{1}{2}(H_x+H_y)$ in this setting, it might be unattainable completely.

As a follow-up, \cite{Kristoufek2015b} studies the issue of $H_{xy}>\frac{1}{2}(H_x+H_y)$ on a theoretical basis in more detail. As it turns out, the answer is almost trivial. The issue is solved through the squared spectrum coherency and its scaling close to the origin. The squared spectrum coherency is defined for two stationary series $\{x_t\}_{t=1}^T$ and $\{y_t\}_{t=1}^T$ with (cross-)spectra $f_{xy}(\omega)$, $f_{x}(\omega)$ and $f_{y}(\omega)$ at frequency $0\le \omega \le \pi$ as 
$$K_{xy}^2(\omega)=\frac{|f_{xy}(\omega)|^2}{f_{x}(\omega)f_{y}(\omega)}$$ for a given frequency $\omega$. Using the definition of the power-law cross-correlations in the frequency domain, we can rewrite the coherency as $$K_{xy}^2(\omega)=\frac{|f_{xy}(\omega)|^2}{f_{x}(\omega)f_{y}(\omega)}\propto \frac{\omega^{2(1-2H_{xy})}}{\omega^{1-2H_x}\omega^{1-2H_y}}=\omega^{2(H_x+H_y-2H_{xy})}.$$ Now note that the squared coherency ranges between 0 and 1 everywhere (in fact even for non-stationary series with their pseudo-spectra). Therefore, it is so restricted for the long-range cross-correlations frequencies as well, i.e. $\omega \rightarrow 0+$. This gives us two feasible and one infeasible possibilities:
\begin{itemize}
\item $H_{xy}=\frac{1}{2}(H_x+H_y) \Rightarrow 2(H_x+H_y-2H_{xy})=0 \Rightarrow \lim_{\omega\rightarrow 0+}{K_{xy}^2(\omega)\propto \text{const.}}$
\item $H_{xy}<\frac{1}{2}(H_x+H_y) \Rightarrow 2(H_x+H_y-2H_{xy})>0 \Rightarrow \lim_{\omega\rightarrow 0+}{K_{xy}^2(\omega)=0}$
\item $H_{xy}>\frac{1}{2}(H_x+H_y) \Rightarrow 2(H_x+H_y-2H_{xy})<0 \Rightarrow \lim_{\omega\rightarrow 0+}{K_{xy}^2(\omega)=+\infty} \Rightarrow \lightning $
\end{itemize}

This implies that $H_{xy}>\frac{1}{2}(H_x+H_y)$ is impossible. Note that this holds for stationary as well as for non-stationary processes (and it can be easily shown for the DCCA fluctuations scaling as well). If the empirical literature reports otherwise, it is due to a bias. This bias might be due to various reasons. First, the standardly used estimators of the bivariate Hurst exponent $H_{xy}$ seem to be biased in general as well as due to short-term dependence bias \citep{Kristoufek2015a} even though the latter should not be the case, at least theoretically \citep{Kristoufek2015}. Second, the estimators are strongly upward biased in presence of heavy tails \citep{Kristoufek2016}, which is usually the case in the financial time series \citep{Cont2001}. Note that the spectrum-based estimators of $H_{xy}$ \citep{Kristoufek2014} are not biased by the heavy tails. And third, there is a finite sample bias as showed in detail in \cite{Kristoufek2015b}. Unfortunately, this bias can be either positive, negative or none depending on the level of correlation between series for scales close to zero. This makes $H_{xy}$ or specifically its comparison with $\frac{1}{2}(H_x+H_y)$ unreliable.

What makes this finding even more alarming is the fact that in the financial econometrics and time series analysis literature, the impossibility of $H_{xy}>\frac{1}{2}(H_x+H_y)$ is taken as an obvious property and it is pretty much a two-liner in \cite{Sela2012} who quickly focus on the $H_{xy}<\frac{1}{2}(H_x+H_y)$ case as the only relevant one for further analysis.

\section{All in vain?}

One might then ask whether the whole research around power-law cross-correlations is in vain and futile. The short answer is ``no'' but it needs further work with more care about theoretical aspects of the topic. As it stands, most of the empirical literature reports either $H_{xy}>\frac{1}{2}(H_x+H_y)$ or $H_{xy}=\frac{1}{2}(H_x+H_y)$. The former is infeasible, i.e. wrong, and the latter is not interesting as it is implied by many different models. In addition, the latter case is simply a reflection of power-law auto-correlations of the separate processes (or at least one of them) and the fact that the processes are pairwise correlated, nothing else is needed for $H_{xy}=\frac{1}{2}(H_x+H_y)$ to hold \citep{Kristoufek2015}. There are two ways how this research branch can be further exploited even without $H_{xy}$ and its sole interpretation -- utilizing the construction of DCCA without needing $H_{xy}$, and focusing on the case of $H_{xy}<\frac{1}{2}(H_x+H_y)$.

\subsection{Scale-specific correlations and regressions}

The DCCA procedure is built on scaling of the bivariate fluctuation function $F^2_{XY}(s)$, which eventually leads to a power-law scaling $F^2_{XY}(s) \propto s^{2H_{xy}}$, in the same way the DFA procedure is based on the fluctuation function $F^2(s)$ scaling. Asymptotically, these can be seen as covariance and variance, respectively, relative to the specific scale $s$, i.e. a scale-specific covariance $F^2_{XY}(s)$ and a scale-specific variance $F^2(s)$. This idea has been further expanded by \cite{Zebende2011} who proposed the DCCA-based correlation coefficient as $$\rho_{DCCA}(s)=\frac{F^2_{XY}(s)}{\sqrt{F_X^2(s)F_Y^2(s)}}$$ where $F_X^2(s)$ and $F_Y^2(s)$ are scale-specific variances of processes $X$ and $Y$. This correlation coefficient has been shown to work well for non-stationary series as well and to outperform the standard Pearson correlation coefficient \citep{Kristoufek2014a}. In addition, its construction is so straightforward and appealing that it is quite easy to construct such correlation coefficients using almost any power-law cross-correlations method \citep{Kristoufek2014b}.

When the scale-dependent correlations are defined, it is only a simple step towards regression. \cite{Kristoufek2015c} introduces a DCCA-based estimator of the scale-dependent $\beta$ coefficient, defined as $$\hat{\beta}^{DCCA}(s)=\frac{F^2_{XY}(s)}{F_X^2(s)}.$$ Compared to $\rho_{DCCA}(s)$, which measures the strength of the relationship, $\hat{\beta}^{DCCA}(s)$ gives the specific effect, i.e. its level, which is much more useful for interpretation of economic and financial relationships where one is usually interested not only in whether the variables are strongly or weakly correlated but what the actual effect of one variable on another is.

The work and insight of \cite{Zebende2011} has thus given a very important alternative utility of the DCCA method (and other time domain $H_{xy}$ in general) and he has shown that the in-between steps of methods can sometimes lead to completely novel views on the topic.

\subsection{Power-law coherency}

As noted by \cite{Sela2012} and \cite{Kristoufek2015b,Kristoufek2017}, only the case of $H_{xy}<\frac{1}{2}(H_x+H_y)$ is an interesting venue as it promises a new class of processes. Returning back to utilizing the squared spectrum coherency, if the two processes are power-law correlated so that $f_{x}(\omega) \propto \omega^{1-2H_x}$ and $f_{y}(\omega) \propto \omega^{1-2H_y}$ close to the origin ($\omega \rightarrow 0+$) and they are power-law cross-correlated so that $|f_{xy}(\omega)| \propto \omega^{1-2H_{xy}}$ close to the origin, we can write
\begin{equation}
\label{eq:coherence}
K_{xy}^2(\omega)\propto \frac{\omega^{2(1-2H_{xy})}}{\omega^{1-2H_x}\omega^{1-2H_y}}=\omega^{-4(H_{xy}-\frac{H_x+H_y}{2})} \equiv \omega ^{-4H_{\rho}} \nonumber
\end{equation}
close to the origin. The power-law coherency can be defined through parameter $H_{\rho}$ as $H_{\rho}=H_{xy}-\frac{H_x+H_y}{2}$.

Recall that the squared spectrum coherency $0\le K_{xy}^2(\omega) \le 1$ for all frequencies $\omega$ which yields only two possible settings for the exponent -- either $H_{\rho}=0$ or $H_{\rho}<0$. $H_{\rho}=0$ gives us $H_{xy}=\frac{H_x+H_y}{2}$ and the coherency goes to a constant for very low frequencies $\omega$. The more interesting situation arises when $H_{\rho}<0$ (resulting in $H_{xy}<\frac{H_x+H_y}{2}$) which implies that the squared coherency goes to zero for low frequencies approaching zero, specifically in the power-law manner (hence power-law coherency). The power-law coherent processes can be correlated at high frequencies but are uncorrelated at low frequencies. From the perspective of financial economics, these processes can be correlated in the short-term but are uncorrelated in the long-term (and that is why \cite{Sela2012} refer to such processes as anti-cointegration). Such processes have potentially huge impact on portfolio construction and risk management as an asset characterized as such would serve as important risk diversifiers from the long-term perspective.

As in detailed shown by \cite{Kristoufek2017}, the power-law coherency can be translated into the time domain easily. Eventually, one arrives at \begin{equation}
\label{eq_PLC}
\rho_{xy}^2(s) \propto \frac{s^{4H_{xy}}}{s^{2H_x}s^{2H_y}} = s^{4(H_{xy}-\frac{H_x-H_y}{2})} \equiv s^{4H_{\rho}} \nonumber
\end{equation}
so that the scaling exponent for both time ($s \rightarrow +\infty$) and frequency ($\omega \rightarrow 0+$) domain power-law coherency is the same. Interestingly, the squared correlation $\rho_{xy}^2(s)$ can be easily represented by the squared DCCA-based correlation coefficient $\rho^2_{DCCA}(s)$ so that both approaches presented in this section and the previous one nicely connect in the end. This parallel view gives another insight into the interpretation of the relationship between $H_{xy}$ and $\frac{1}{2}(H_x+H_y)$, specifically two interesting cases. When $H_{xy}=\frac{H_x+H_y}{2}$ and $\rho_{xy}^2(s) \approx 1$ for $s \rightarrow +\infty$, we have possible cointegration, i.e. the variables are not necessarily connected in the short-term but are tightly connected in the long-term. And when $H_{xy}<\frac{H_x+H_y}{2}$, we have anti-cointegration, i.e. the variables are not related in the long-term but are possibly connected in the short-term.

\section{Discussion}

The whole issue and suggested solutions presented above point mainly to a general problem of interdisciplinary research (here specifically econophysics) -- communities do not interact enough. The bivariate breakthrough in the sense of power-law cross-correlations came in 2008 \citep{Podobnik2008} but the interpretation of the bivariate cross-correlations parameter was already obviously problematic. Already in 2009, \cite{Sela2009} discussed the multivariate fractionally integrated processes (power-law cross-correlated in the econophysics language) and took as a given that the bivariate cross-correlation parameter is practically forced by the auto-correlation properties of the separate processes. This was only further extended in \cite{Sela2012} but the crucial result has been there since 2009. Moreover, the topic of bivariate dependence in the long-range correlations setting was not new and it had been discussed in the financial econometrics literature much earlier \citep{Lobato1997,Lobato1999} and already \cite{Lobato1999} discusses the connection between the separate and bivariate fractional integration parameters (equivalent to the long-range dependence in the financial econometrics language). Only it has not come to the econophysics community and it had to be re-discovered first by \cite{Podobnik2008}, who introduced the DCCA estimator of the $H_{xy}$ parameter, and \cite{Kristoufek2015b}, who showed that $H_{xy}>\frac{H_x+H_y}{2}$ is impossible and only $H_{xy}<\frac{H_x+H_y}{2}$ is theoretically and practically appealing, and translated the framework into the more standard (for econophysics) time domain language.

However, both directions of communication can be beneficial. Financial econometrics focuses a lot on theoretical properties of estimators as well as their restrictions and assumptions. As shown in the text above, the power-law cross-correlations literature has missed a lot of it since the very beginning. The econophysics community can learn from it and see that quite often, there is a need for theoretical background before jumping into empirical avenues. From the other side, financial econometrics often focuses on assumptions and restrictions too much, missing a big picture and possible usefulness of various methods. The time domain approaches that are so characteristic for the econophysics field are usually omitted in the financial econometrics literature because it is more complicated to show their statistical properties compared to the frequency domain approaches. However, when the methods from both domains are compared in a horse race, the time domain methods often prevail (compare the results in \cite{Kristoufek2014,Kristoufek2015b,Kristoufek2016,Kristoufek2017}).

With increasing computational power, the non-existence (or negligence) of the asymptotic properties for the econophysics methods is becoming less of a problem as the properties can be simulated even for very long time series with various dynamic properties. As it turns out, many of these econophysics methods can compete and even beat the frequency domain methods, which are popular in the economics and econometrics mainstream, even for very long time series which would standardly be considered as a good enough approximation of infinity, i.e. asymptotics. Econophysics has always boasted of being data-driven, empirically based science discipline, which has certainly led to many breakthroughs. However, and this is specifically true for the last couple of years, the econophysics literature has been flooded with empirical papers that are simple analyses of the ``choose the method, input the data, list the results'' type without much effort of results interpretation. For econophysics to be treated with more respect by the mainstream economics and finance communities, the concrete practical implications and applications, such as specific policy suggestions, trading strategies, portfolio methods, and similar, need to be presented. The same issue has been the case for the power-law cross-correlations since the very beginning outlined by \cite{Podobnik2008}. There has been no interpretation or practical implications of the bivariate Hurst exponent $H_{xy}$. It has been standardly stated that the two series are ``power-law cross-correlated'' or ``cross-persistent'' with no hint of what it actually implies (in a practical sense) for the series dynamics.

In this text, it has been shown that the power-law cross-correlations setting is inherently problematic and the bivariate Hurst exponent alone does not give any information about the relationship between two analyzed series. Only a comparison of $H_{xy}$ with its separate counterparts gives any information. However, unless the series are pairwise independent (or at least uncorrelated) or they are the unique case of anti-cointegration, it will automatically hold that $H_{xy}=\frac{H_x+H_y}{2}$ which, unfortunately, covers a wide range of very different possible relationships between the series so that its informative value is again miniature. Fortunately, there are at least two ways (recalled in this text) utilizing partial steps and results of the power-law cross-correlations setting -- scale-specific correlations and regressions, and power-law coherency. Nevertheless, the challenges for these two approaches remain the same and should not be overshadowed by purely empirical studies -- to show practical utility of the methods. For each of these, the utility seems at hand -- portfolio construction. This is the case both for the scale-specific correlations and regression as these can be used e.g. as standard correlation matrices or $\beta$-parameters in the capital asset pricing model setting, and for the anti-cointegration case which promises high quality long-term diversifiers. Only then will we be allowed to say that power-law correlations have been contributive and useful.

\section*{Acknowledgements}
 
Ladislav Kristoufek gratefully acknowledges financial support of the Czech Science Foundation (project 17-12386Y).


\section*{References}
\bibliography{Biblio}
\bibliographystyle{chicago}

\end{document}